\documentclass[%
 reprint,
 amsmath,amssymb,
 aps,
 prb,
]{revtex4-2}

\usepackage{graphicx}
\usepackage{dcolumn}
\usepackage{bm}
\usepackage{hyperref}

\begin{document}

\preprint{APS/123-QED}

\title{Temperature dependent coherence properties of NV ensemble in diamond up to 600K}

\author{Shengran Lin}
\affiliation{Key Laboratory of Strongly-Coupled Matter Physics, Chinese Academy of Sciences, Hefei National Laboratory for Physical Science at Microscale, and Department of Physics, University of Science and Technology of China. Hefei, Anhui, 230026, P. R. China.}

\author{Changfeng Weng}
\affiliation{Key Laboratory of Strongly-Coupled Matter Physics, Chinese Academy of Sciences, Hefei National Laboratory for Physical Science at Microscale, and Department of Physics, University of Science and Technology of China. Hefei, Anhui, 230026, P. R. China.}

\author{Yuanjie Yang}
\affiliation{Key Laboratory of Strongly-Coupled Matter Physics, Chinese Academy of Sciences, Hefei National Laboratory for Physical Science at Microscale, and Department of Physics, University of Science and Technology of China. Hefei, Anhui, 230026, P. R. China.}

\author{Jiaxin Zhao}
\affiliation{Key Laboratory of Strongly-Coupled Matter Physics, Chinese Academy of Sciences, Hefei National Laboratory for Physical Science at Microscale, and Department of Physics, University of Science and Technology of China. Hefei, Anhui, 230026, P. R. China.}

\author{Yuhang Guo}
\affiliation{Key Laboratory of Strongly-Coupled Matter Physics, Chinese Academy of Sciences, Hefei National Laboratory for Physical Science at Microscale, and Department of Physics, University of Science and Technology of China. Hefei, Anhui, 230026, P. R. China.}

\author{Jian Zhang}
\affiliation{Key Laboratory of Strongly-Coupled Matter Physics, Chinese Academy of Sciences, Hefei National Laboratory for Physical Science at Microscale, and Department of Physics, University of Science and Technology of China. Hefei, Anhui, 230026, P. R. China.}

\author{Liren Lou}
\affiliation{Key Laboratory of Strongly-Coupled Matter Physics, Chinese Academy of Sciences, Hefei National Laboratory for Physical Science at Microscale, and Department of Physics, University of Science and Technology of China. Hefei, Anhui, 230026, P. R. China.}

\author{Wei Zhu}
\affiliation{Key Laboratory of Strongly-Coupled Matter Physics, Chinese Academy of Sciences, Hefei National Laboratory for Physical Science at Microscale, and Department of Physics, University of Science and Technology of China. Hefei, Anhui, 230026, P. R. China.}
\author{Guanzhong Wang}

\email[]{gzwang@ustc.edu.cn}
\affiliation{Key Laboratory of Strongly-Coupled Matter Physics, Chinese Academy of Sciences, Hefei National Laboratory for Physical Science at Microscale, and Department of Physics, University of Science and Technology of China. Hefei, Anhui, 230026, P. R. China.}

\date{\today}

\begin{abstract}
Nitrogen-vacancy (NV) center in diamond is an ideal candidate for  quantum sensors because of its excellent optical and coherence property. However, previous studies are usually conducted at low or room temperature. The lack of full knowledge of coherence properties of the NV center at high temperature limits NV's further applications. Here, we systematically explore the coherence properties of NV center ensemble at temperature from 300\,K to 600\,K. Coherence time $T_2$ decreases rapidly from $184\ \mu s$ at 300\,K to $30\ \mu s$ at 600\,K, which is attributed to the interaction with paramagnetic impurities. Single-quantum and double-quantum relaxation rates show an obvious temperature-dependent behavior as well, and both of them are  dominated by the two phonon Raman process. While the inhomogeneous dephasing time $T_2^*$ and thermal echo decoherence time $T_{TE}$ remain almost unchanged as temperature rises. Since $T_{TE}$ changed slightly as temperature rises, a thermal-echo-based thermometer is demonstrated to have a sensitivity of $41\ mK/\sqrt{Hz}$ at 450\,K. These findings will help to pave the way toward NV-based high-temperature sensing, as well as to have a more comprehensive understanding of the origin of decoherence in the solid-state qubit.

\end{abstract}

\maketitle


\section{Introduction}

Negatively charged nitrogen-vacancy (NV) center in diamond has attracted much attentions for their excellent properties at room temperature, such as long coherence time, optically initialize and read out. These qualities make them an appealing candidate for applications ranging from quantum information processing  \cite{1}, quantum computing  \cite{2} to quantum sensing  \cite{3,4,5,6}. Moreover, the high thermal conductivity and stability of diamond makes NV-based devices be able to perform at high temperature \cite{7,8}, which are of great potential application for the exploration of temperature-related phenomena, like magnetic phase transition \cite{9}, and thermoelectric effects \cite{10}. 

It is pointed out that NV center can be coherently manipulated and hold a long inhomogeneous dephasing time at temperature up to 625\,K \cite{7}. By initializing and reading out at room temperature, one can manipulate the NV center at temperature up to 1000\,K \cite{8}. However, prior studies of NV center at high temperature are mainly focused on relaxation time $T_1$ and inhomogeneous dephasing time $T_2^*$  \cite{7,8,12}, and much less is known about the coherence time $T_2$, which is one of the most important parameters of a qubit. It not only sets the sensitivity of the corresponding sensor, but also constrains the maximum number of gate operations in quantum computing or quantum information processing. Moreover, it should take consideration to the double quantum (DQ) relaxation reside in the three level system (Fig. \ref{fig:1}(a)) of NV center, for which will have a significant impact on the decoherence of NV center, especially for those near surface  \cite{13}. Further applications of NV center at high temperature thus require a fully understanding of the coherence properties. In addition, most of the studies are using single NV center, while the utilization of NV center ensemble would dramatically improve the sensitivity of NV-based sensor ($\eta \propto 1/\sqrt{n}$).

In this work, we systematically study the coherence property of NV center ensemble at temperature from 300\,K to 600\,K. The spin echo pulse experiments show that the decoherence time $T_2$ decreases rapidly as temperature rises, while the inhomogeneous dephasing time $T_2^*$ and thermal-echo (TE) dephasing time $T_{TE}$  remain almost unchanged, with the optically detected magnetic resonance (ODMR) contrast decreased. The rate of relaxation process, involving single quantum (SQ) and DQ transitions, both increase repidly as temperature rises. We also demonstrated a high temperature thermometer based on NV center ensemble at around 450\,K. These findings will help to improve the performance of NV-based devices at high temperature.

\section{Experiment}

\begin{figure} 
	\centering
	\includegraphics{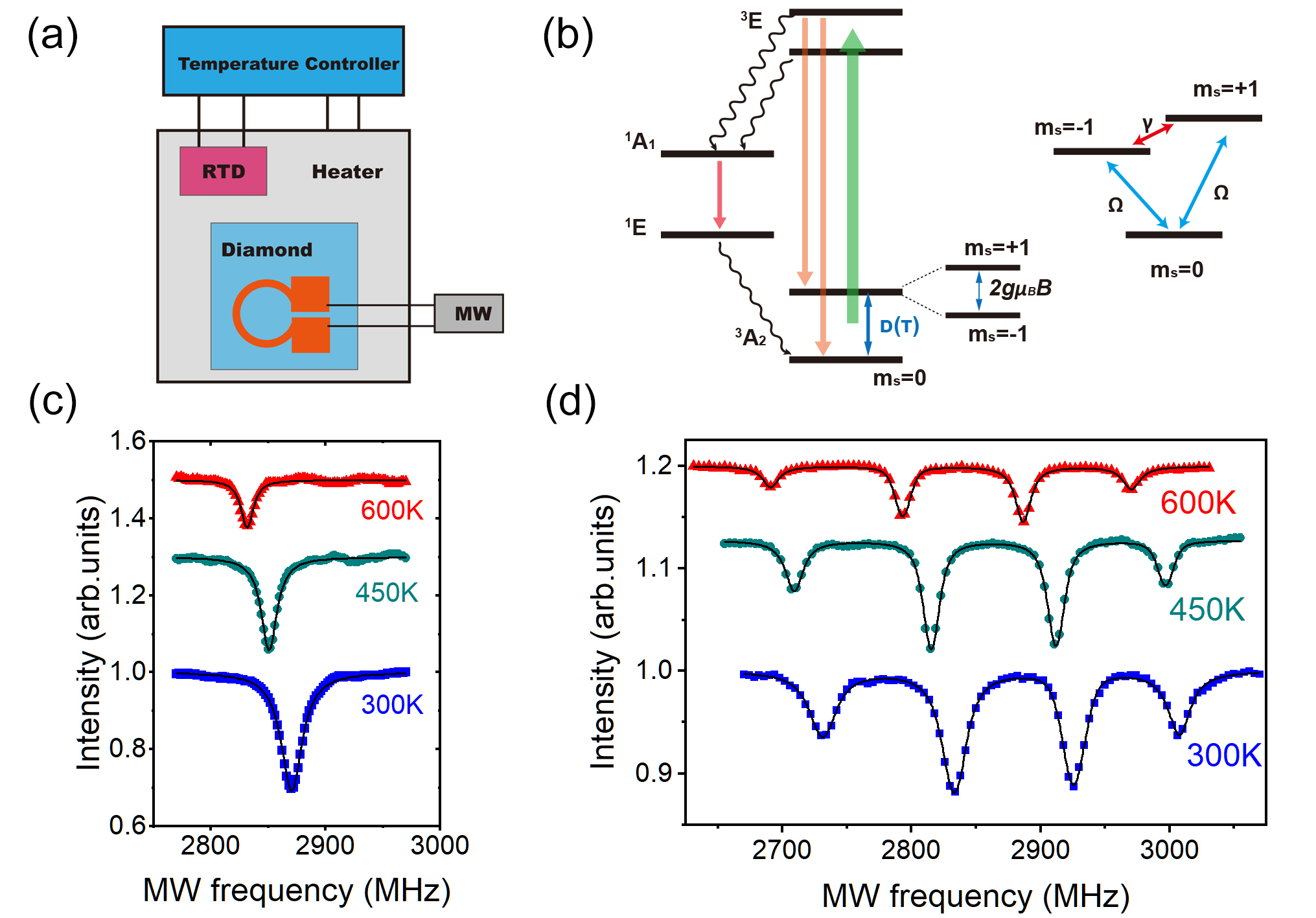}
	\caption{(a) Schematic showing the experiment arrangement of the sample and the heating part. The diamond sample is adhered on a resistance heating plate with the temperature detected by a RTD and controled with a temperature controller. And a ring-shape antena is set on the sample to deliver MW field. (b) The electronic energy level diagram of NV center, the degenerated sublevel $|m_s=\pm1\rangle$ is splited with magnatic field applied. The straight lines represent optical transitions and the snaked lines represent non-radiative decay. (c) Zero field CW ODMR spectra at three tempratures, 300\,K, 450\,K, 600\,K(From the bottom up).  (d) CW ODMR at three temperatures with a 50G magnetic field applied.}
	\label{fig:1}
\end{figure}

A diamond plate sample grown by plasma-enhanced CVD method is used in the experiment ($[N]\approx125 \,ppb$, $[NV^- ]\approx2\,ppb$ ($1\,ppb=1.76\times10^{17}/cm^3$)). The concentration of nitrogen was determined by Electron Paramagnetic Resonance (EPR,JES-FA200) and the concentration of NV center was estimated by comparing the photoluminescence (PL) intensity of the sample with that of a single NV center. To coherently manipulate the NV center, a home-built confocal microscope (CFM) equipped with a microwave (MW) system was used. The NV center was excited by a 532\,nm laser and the resulting fluorescence was collected and directed to a single photon counter (SPCM-AQRH-W4). To avoid the heating effect of laser beam, the power of laser beam ahead of the objective lens was kept at 0.2\,mW in all the subsequent measurements. The MW was delivered to the NV center through the ring-shape coplanar waveguide on the diamond plate (Fig. \ref{fig:1}(a)). More detail of the setup can be found in our previous report \cite{14}. To study the temperature-dependent coherence property of NV center, a metal ceramic heater (HT24S, Thorlabs) attached to the diamond and a resistive temperature detector (TH100PT, Thorlabs) fixed closely  to the diamond were used to control the temperature of the diamond (Fig. \ref{fig:1}(b)).

As shown in Fig. \ref{fig:1}(c), the continuous-wave (CW) ODMR spectra of NV center exhibit a resonance peak, corresponding to the zero-field splitting (ZFS) between the $m_s=\pm1$ and $m_s=0$ spin sub-levels of NV center (denote as D, see Fig. \ref{fig:1}(a)). As temperature rises, both D’s value and ODMR contrast decrease, which is consistent with the previous researches  \cite{7,15} . The decrease of D value is ascribed to the thermal expanding and electron-phonon interaction  \cite{16}, which is the bases of NV thermometer. As for the ODMR contrast, it was ascribed to the shorter lifetime of  $m_s=\pm1$ and $m_s=0$ states at high temperature  \cite{7}. According to our results and that reported in the literatures  \cite{7,16,17}, the temperature dependence of D varies little among diamond samples, thus we used the D(T) relationship reported in ref  \cite{7} to determine the temperature in the subsequent experiments, which was also confirmed by the temperature dependence of the diamond Raman shift  \cite{19}.

We then conducted pulse ODMR experiments to extract the coherence times of NV center at various temperatures. To lift the degeneracy of the $m_s=\pm1$ spin sublevels and to spectrally distinguish the NV center with different crystallographic orientations, a magnetic field of about 50\,G was applied by using a high-temperature magnet ($Sm_2 Co_{17}$), with the magnetic field direction adjusted to be parallel to one of the $\langle111\rangle$ direction of the diamond. The corresponding CW spectrum shows four resonances (Fig. \ref{fig:1}(d)), and we chose the two most outer resonances, which were belong to the NV center with the axes parallel to the magnetic field direction, to be employed in the subsequent experiment. Then we applied various pulse sequences to obtain the relevant coherence times.

\section{Result and discussion}

\subsection{Ramsey and thermal-echo measurement}

\begin{figure}
	\centering
	\includegraphics{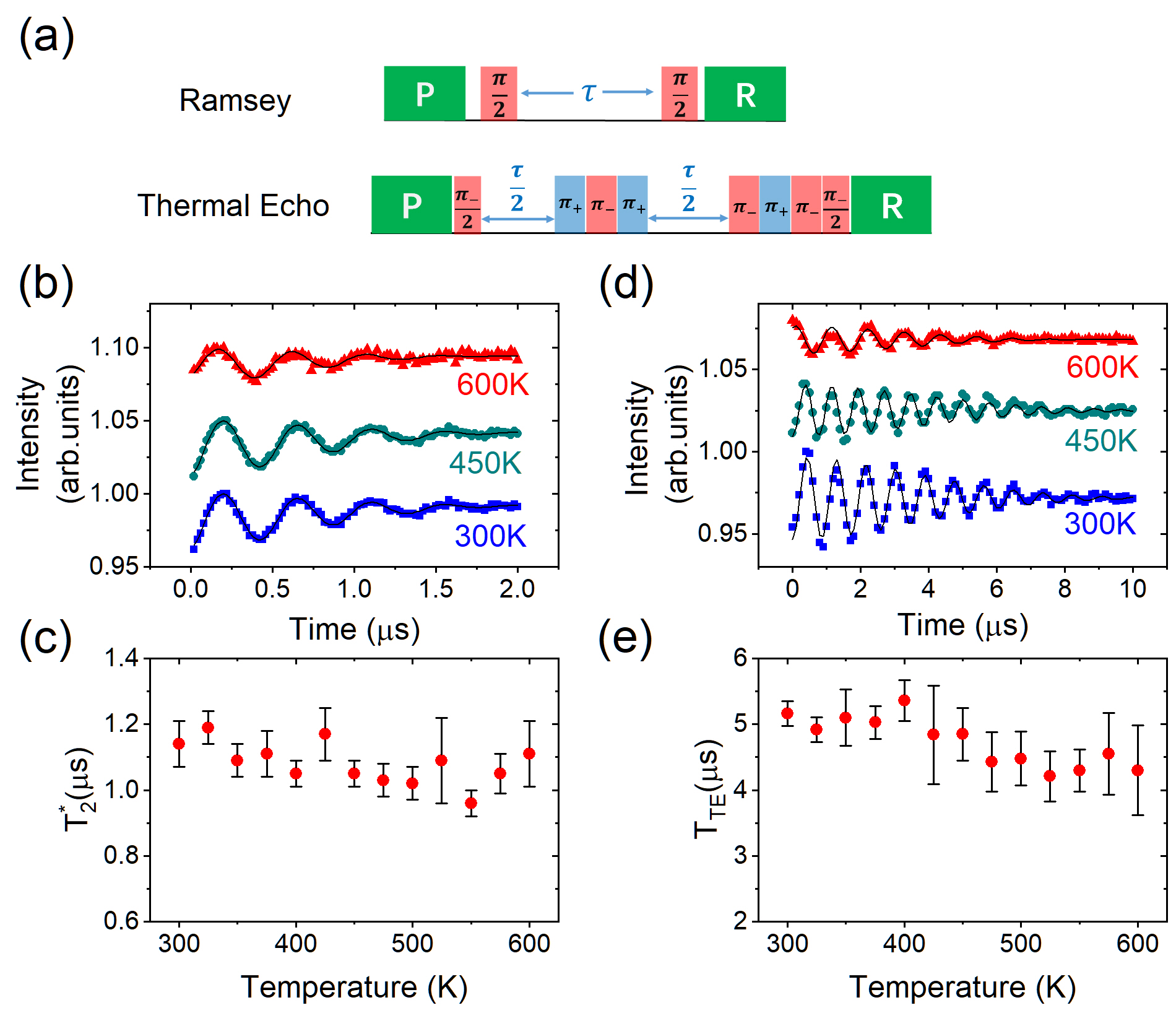}
	\caption{Ramsey and Thermal echo measurement.(a) Pulse sequence for Ramsey and Thermal echo respectively. Upper panel: Ramsey, lower panel: Thermal echo. Minus (-) denotes the transition $|0\rangle \leftrightarrow |-1\rangle$  , and plus (+) denotes the transition $|0\rangle \leftrightarrow |+1\rangle$. (b) Ramsey fringes at 300K,450K and 600K (from the bottom up). (c) Inhomogeneous dephasing time $T_2^*$ from 300\,K to 600\,K.(d) Thermal echo curves at 300\,K, 450\,K and 600\,K (from the bottom up). (e) Thermal echo coherence time $T_{TE}$ from 300\,K to 600\,K. Error bars in (c) and (e) represent fitting error. }
	\label{fig:2}
\end{figure}

Ramsey and TE  \cite{6} pulse sequences are extensively utilized for quantum sensing, in which the inhomogeneous dephasing time $T_2^*$ and TE dephasing time $T_{TE}$ are the crucial parameters relating to the sensing sensitivity. The pulse sequence for Ramsey measurement is shown in Fig. \ref{fig:2}(a). The two laser pulses at the beginning and the end of the sequence are used to polarize and read out the spin state of NV center respectively. To coherently manipulate the NV center, resonant MW was applied. The first MW $\pi/2$ pulse is used to transform the NV $|0\rangle$ state into the superposition state $|\phi\rangle=1/\sqrt{2} (|0\rangle+|-1\rangle) $or other equilibrium states, and the second MW $\pi/2$ pulse is used to transform the accumulated phase information into population difference. By varying the interval between two MV pulses, we will obtain the Ramsey fringe and the corresponding inhomogeneous dephasing time. 

In the experiment, we carried out Ramsey measurement at temperature from 300\,K to 600\,K (Fig. \ref{fig:2}(b)). The oscillation of the curves was ascribed to the MW frequency detuning from the resonance and the interaction with the $^{14}N$ nuclear spin. The contrast decreases as temperature rises, while the dephasing time $T_2^*$ keeps unchanged (Fig. \ref{fig:2}(c)). The pulse sequence for TE measurement is shown in Fig. \ref{fig:2}(a). For the TE measurement, both $|0\rangle \leftrightarrow |-1\rangle$ and $|0\rangle \leftrightarrow |+1\rangle$ transitions were utilized to reduce the dephasing effect from stray magnetic field, thus a longer dephasing time can be obtained comparing to the $T_2^*$ derived from Ramsey measurement. Fig. \ref{fig:2}(d) shows the TE curves at three temperatures (300\,K, 450\,K, 600\,K) and Fig.\ref{fig:2}(e) shows that $T_{TE}$ change slightly from 300\,K to 600\,K.

For the diamond sample we used in this work, $^{13}C$ nuclear spin is the dominant dephasing source responsible to $T_2^*$. Since the experiment temperature is well above the polarization temperature of $^{13}C$ nuclear spin ($k_B T\gg \mu_I B$), $^{13}C$ nuclear spins are considered to have an equal population at every sub-level. Moreover, a recent research shows that the hyperfine interaction between NV electron spin and $^{13}C$ nuclear spin keeps nearly unchanged at temperature up to 700\,K  \cite{20}. Thus, the dephasing effect from the $^{13}C$ nuclear spin bath was almost temperature-independent, leading to a temperature independent $T_2^*$. The robust behavior of $T_{TE}$ should be related to the same mechanism. These results indicated that the Ramsey-based or TE-based sensor can operate reliably at temperature up to 600\,K.

\subsection{Relaxation and spin echo measurement}

\subsubsection{Relaxation}

\begin{figure}
	\centering
	\includegraphics{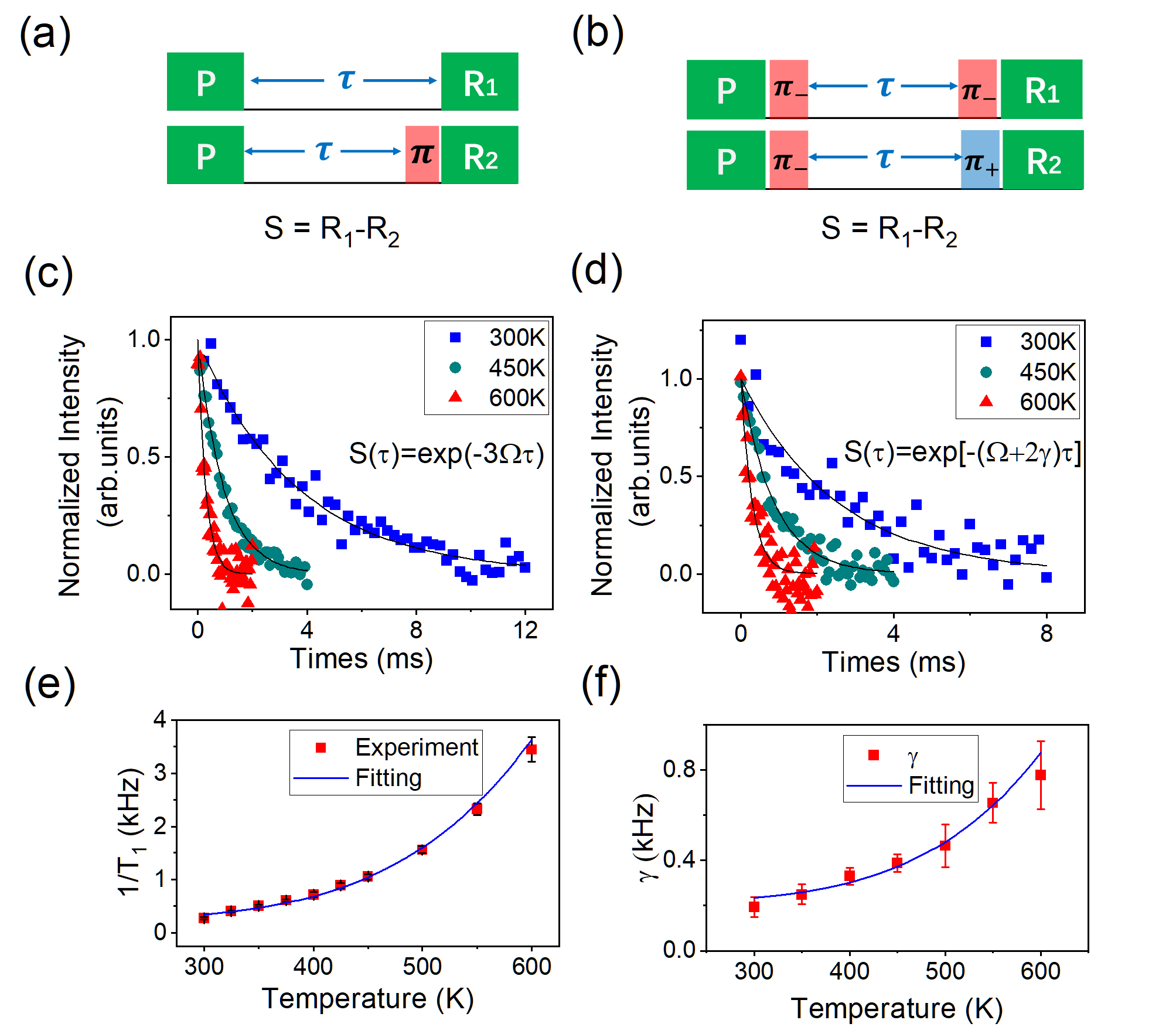}
	\caption{DQ and SQ relaxation experiment. (a,b) Measurement sequences to extract the relaxation rate $\Omega$ and $\gamma$. $\pi_-$ corresponds to the resonance $|0\rangle \leftrightarrow |-1\rangle$, and $\pi_-$ corresponds to the resonance $|0\rangle \leftrightarrow |+1\rangle$. (c) Relaxation curves acquired by using sequence in (a) at 300\,K, 450\,K, 600\,K respectively. The data are fitted by function $exp(-3\Omega \tau)$). (d) Relaxation curves acquired by using sequence in (b) at 300\,K, 450\,K, 600\,K respectively. The data are fitted by function $exp[-(\Omega+2\gamma)\tau]$. (e) Temperature dependence of longitudinal relaxation rate $1/T_1$  ($3\Omega$). Red points are experiment data, and the solid line is the fitting curve ($A_{01} T^5+B_{01}$). (f) Temperature dependence of double quantum relaxation rate $\gamma$. Red points are experimental data, and the solid line is the fitting curve ($A_{\gamma} T^5+B_{\gamma}$). Error bars correspond to fitting error.}
	\label{fig:3}
\end{figure}

We then conducted relaxation and Hahn echo measurements to determine the relaxation rate and coherence time $T_2$. Since the ground state of NV center is a three-level system (Fig. \ref{fig:1}(a)), the DQ transition ($\Delta m=\pm2$) will make contribution to the relaxation process, which will affect the SQ coherence. In this work, SQ transition denoted the transition between $|m_2=0\rangle$ and $|m_s=-1\rangle$ states, and DQ transition denoted the transition between $|m_s=-1\rangle$ and $|m_s=+1\rangle $ states. We used the sequences in Fig. \ref{fig:3}(a) and Fig. \ref{fig:3}(b) to obtain the single quantum (SQ) transition rate $\Omega$ and DQ transition rate $\gamma$. Generally, $\Omega$ is related to the ordinary $T_1$ with relationship $3\Omega = 1/T_1$ \cite{13}. Fig. \ref{fig:3}(c) (Fig. \ref{fig:3}(d)) shows the relaxation curves acquired by using sequence in Fig. \ref{fig:3}(a) (Fig. \ref{fig:3}(b)) at three representative temperatures (300\,K, 450\,K, 600\,K). The transition rates $\Omega$ and $\gamma$ are obtained by fitting the data in Fig. \ref{fig:3}(c) and Fig. \ref{fig:3}(d) with function $exp(-3\Omega \tau)$ and $exp[-(\Omega+2\gamma)]$ \cite{13}($\tau$ is the free decay time) respectively. both the SQ relaxation and DQ relaxation become faster as temperature rises. At room temperature, we find that $\gamma/\Omega\,\approx \, 2.1$, indicating that the DQ transition should not be neglected in the measurement of coherence time. Fig. \ref{fig:3}(e) shows the longitudinal relaxation rate $1/T_1$  ($3\Omega$) increase rapidly as temperature rise, which are fitted quite well with function $AT^5+B$. This temperature dependence of relation rate indicate that the SQ relaxation is dominated by a two phonon Raman process  \cite{21}, which agrees well with previous studies  \cite{8,12}. The values of the fitting are $A_{01}=4.4(2)\times10^{-11} \, K^{-5} s^{-1}$, $B_{01}=237(20) \, s^{-1}$, indicating that $A_{\Omega}=A_{01}/3=1.46(7)\times10^{-11} \, K^{-5} s^{-1}$,$B_{\Omega}=B_{01}/3=79(7)  s^{-1}$. Fig. \ref{fig:3}(f) shows the DQ transition rate $\gamma$ as a function of temperature. From Walker’s result  \cite{21}, the DQ relaxation should be dominated by the two Raman process as well and the transition rate depends little on the energy difference between the transition levels. Fig. \ref{fig:3}(f) shows that the $T^5$ fit is in excellent agreement with experimental data. The values of the fitting are $A_{\gamma }=0.85(9)\times10^{-11}\,  K^{-5} s^{-1}$, $B_{\gamma }=215(20) \, s^{-1}$. The discrepancy of $A_{\gamma }$ and $A_{\Omega}$ may be ascribed to the fact that the transition can also be induced by fluctuation field with frequency matching the energy difference between two level (magnetic field for SQ transition and electric field for DQ transition). For near-surface NV center, the surface-related electric field will make large contribution to the DQ transition rate $\gamma $, and the understanding of temperature dependence of $\gamma $ can help us to identify the originals of electric field noise.

\subsubsection{Hahn Echo measurement}

\begin{figure}
	\centering
	\includegraphics{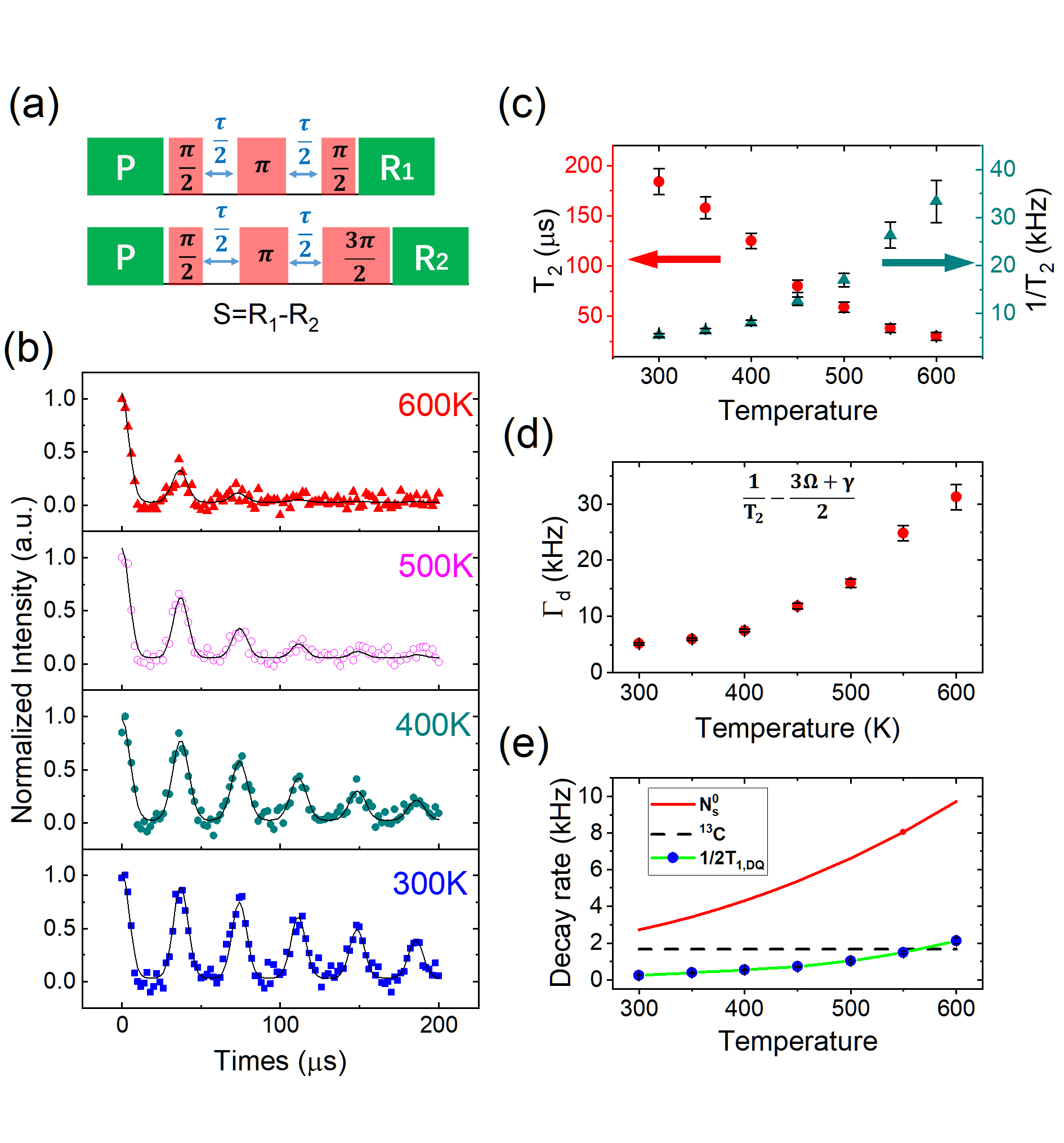}
	\caption{Hahn echo measurement. (a) Pulse sequences for Hahn echo measurement. (b) Hahn echo curves at four temperatures, 300\,K, 400\,K, 500\,K, 600\,K (from the bottom up). The experiment data are fitted by function $exp[-(\tau/T_2 )^p ]\times\sum_i exp[-(\tau-i\times T_R)^2/T_w^2]$, with $p$, $T_R$, $T_w$ being the fit parameters.rs. (c) The temperature dependence of coherence time $T_2$ (decoherence rate $1/T_2$). (d) The temperature dependence of pure dephasing rate, $\Gamma_d=1/T_2 -(3\Omega+\gamma)/2$. (e)The temperature dependence of decochrence effect from relaxation (blue dots), P1 center (red solid line), $^{13}C$ nuclear spin (black dashed line). The contribution from P1 center and $^{13}C$ nuclear spin are estimated from Eq. (2) and Eq. (3), while the contribution from relaxation is determined from experiment. The error bars correspond to fitting errors.}
	\label{fig:4}
\end{figure}

In Hahn echo measurement, the utilization of a $\pi$ pulse in the middle of the free evolution duration can refocus the dephasing caused by stray field, leading to a longer dephasing time. To eliminate the common mode noise (relaxation of off-resonance NV- center), we used the sequence shown in Fig. \ref{fig:4}(a) to perform the measurement. Fig. \ref{fig:4}(b) are the Hahn echo curves at four representative temperatures (300\,K, 400\,K, 500\,K, 600\,K). The collapse and revival of the signal is ascribed to the interaction with the nuclear spin of $^{13}C$ in diamond, with the revival period to be $T_R=2000/(\gamma_{^{13}C} B)$ ($\gamma_{^{13}C} =1.071$ is the gyromagnetic ratio of nuclear $^{13}C$ and B is the magnitude of the applied magnetic field). To obtain $T_2$, we fitted the data by function $exp[-(\tau/T_2 )^p ]\times\sum_i exp[-(\tau-i\times T_R)^2/T_w^2]$, with $p$, $T_R$, $T_w$ being the fit parameters. Unlike the robust behavior of $T_2^*$ and $T_{TE}$, $T_2$ decreased almost linearly from $184\, \mu s$ at 300\,K to $30\,\mu s$ at 600\,K (Fig. \ref{fig:4}(c)). Once the sample cooled down, $T_2$ can recover to the original value (Data not showed here).

For a three-level system, the constraints of $T_2$ is  \cite{13} 
\begin{equation}
\frac{1}{T_2} = \Gamma_d+\frac{3\Omega+\gamma}{2}
\end{equation}

Where $\Gamma_d$ we refer to the pure dephasing rate, and $(3\Omega+\gamma)/2$ is the contribution from lifetime broadening, in which the three-level nature of NV have been considered. Our results have shown that $\Omega$ and $\gamma$ are both temperature-dependent, thus the lifetime broadening will make contribution to the decrease of $T_2$ as temperature rises. To identify the temperature behavior of pure dephasing, we subtracted the corrected relaxation rate ($(3\Omega+\gamma)/2$) from the measured decoherence rate ( $1/T_2$  ). Fig. \ref{fig:4}(d) shows that the pure dephasing rate $\Gamma_d$ increases as temperature rises. For NV center in diamond, the main dephasing mechanism is the dipolar interaction with the surrounding spins, whose flip will cause fluctuations of the local field at the NV center. The spins can flip by two means: spin-lattice (SL) relaxation and spin-spin (SS) relaxation. And the corresponding contributions to the dephasing are given by  \cite{22}
\begin{equation}
\frac{1}{T_{SL}}=\frac{1}{1.4} (\frac{2.53\mu_0 g_e g_A \beta_e \beta_A}{4\pi \hbar} \frac{c_A}{T_1^A})^{1/2}
\end{equation}
\begin{equation}
\frac{1}{T_{SS}}=\frac{0.37\mu_0(g_e\beta_e)^{1/2}(g_A \beta_A)^{3/2}[S(S+1)]^{1/4}}{2} c_A
\end{equation}

In which A denotes the surrounding spins, $\mu_0=4\pi\times10^{-7}\,T^2 J^{-1} m^3$ is the permeability of vacuum, $g_e=2$ is the g-factor of NV center, $g_A$ is the g-factor of the A spins, $\beta_e$ is the Bohr magneton, $\beta_A$ is Bohr magneton or nuclear magneton, depending on the spin type being electron spin or nuclear spin, $c_A$ is the number of A spins per unit volume, $T_1^A$ is the longitudinal relaxation time of A spins.  It’s apparently that  $1/T_{SS}$ is temperature-independent. For the diamond sample used in this work, the main paramagnetic impurities are the electron spin of substitutional nitrogen ($[N_s^0]=0.125\, ppm$, P1 center) and the nuclear spin of $^{13}C$ ($[^{13}    C]\approx 1.1\%$). Schematically, the pure dephasing rate of NV center can be expressed as  \cite{23}

\begin{equation}
\Gamma_d = \frac{1}{T_2,\{N_s^0\}}+\frac{1}{T_2,\{^{13}C\}}+\frac{1}{T_2,\{others\}}
\end{equation}

For $^{13}C$ nuclear spin, the longitudinal relaxation time is fairly long, thus the contribution from spin-lattice relaxation can be neglected, $\frac{1}{T_2,\{^{13}C\}}\approx \frac{1}{T_{SS},\{^{13}C\}}$.  For P1 center, the longitudinal relaxation is dominated by spin-orbit phonon-induced tunneling at high temperature, with $1/T_1^N=A_N T+B_N T^5$(T is the temperature, A and B are constants) \cite{24}. In this article, we used $A_N=5\times10^{-5}\,  K^{-1} s^{-1}$, $B_N=1.1\times10^{-10}\, K^{-5} s^{-1}$ to calculate the $1/T_{LL},\{N_s^0\}$. Fig 4(e) illustrates the decoherence effect from $^{13}C$ nuclear spin and P1 center respectively. It is apparent that P1 center is the dominating decoherence source, with the decoherence rate increasing rapidly as temperature rises. While the decoherence contribution from $^{13}C$ nuclear keep unchanged. The contribution from life broadening of NV itself is also showed, but it make little contribution comparing to other effects. However, the decoherence effect we considered here can not fully represent the decrease of $T_2$ as temperature rise. We ascribed this discrepancy to other paramagnetic impurities we did not consider here, like NVH  \cite{Khan2013}, vacancy cluster \cite{Hounsome2006}, and $SiV^-$ \cite{Feng1993} et al, which have been found in CVD diamond. 

\section{High temperature thermometer}

\begin{figure}[htp]
	\centering
	\includegraphics{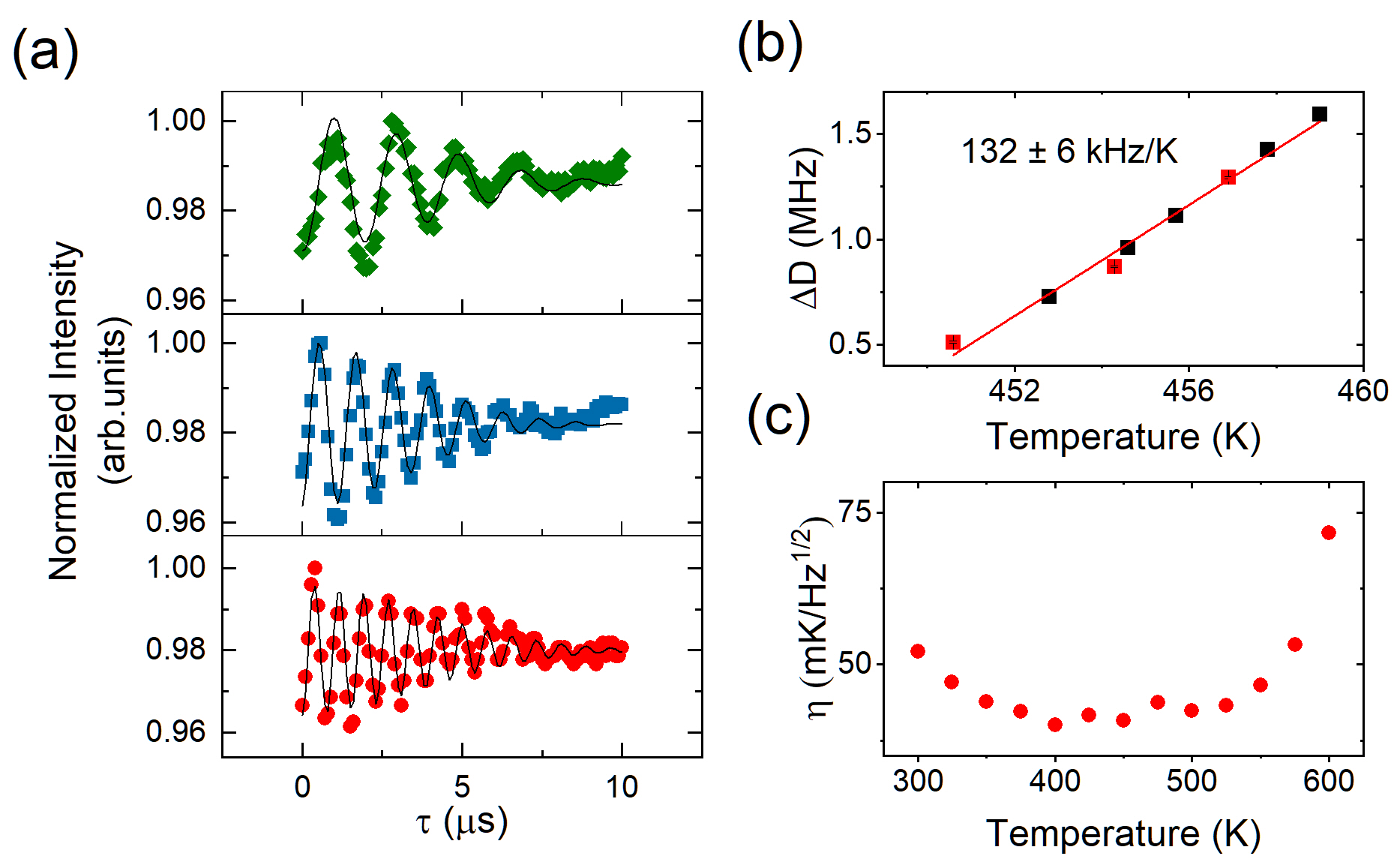}
	\caption{Thermal echo based thermometer. (a) Thermal echo curves at different temperatures, with the applied microwave frequencies fixed. Solid lines are fitting curves. (b) Temperature dependence of thermal echo curve oscillation frequency. The red solid line is a liner fitting, with a slope of $132\pm6\, kHz/K$. (c) The thermal sensitivity obtained from Eq.\,(5), with all parameter determined from experiment.}
	\label{fig:5}
\end{figure}

Since NV center has their TE dephasing time invariant at temperature up to 600\,K, we demonstrate a NV ensemble based high temperature thermometer by using the TE technology. The prototype thermometer was tested at temperature region around 450\,K. In the experiment, the MW frequencies remained unchanged. And the oscillation frequency of thermal-echo curve ($f=(\omega_-+\omega_+)/2-D$, in which $\omega_-$ and $\omega_+$ are applied MW frequencies) was used to extract the D value. Fig. \ref{fig:5}(a) are three representative thermal-echo curves measured at different temperatures, which shows oscillation with frequency of $1295\, kHz$, $870\, kHz$ and $512\, kHz$ respectively. Fig5. (b) shows that the oscillation frequencies are linearly dependent on temperature with a slope of $132\pm6 \,kHz/K$, which is consistent with the temperature dependence of D  \cite{7}. To evaluate the sensitivity of our TE based thermometer, we use the equation \cite{6}

\begin{equation}
\eta = \sqrt{\frac{2(p_0+p_1)}{p_0-p_1}^2} \frac{1}{2\pi \frac{dD}{dT} exp(-(\frac{t}{T_{TE}})^m)\sqrt{t}}
\end{equation}

 in which $p_0$ and $p_1$ are photo count per measurement shot of the NV ensemble in bright and dark state respectively. $T_{TE}$ is the TE dephasing time, D is the zero-field splitting, and the sensitivity corresponds to the maximum value of $exp(-(\frac{t}{T_{TE}})^m)\sqrt{t}$. We determined the sensitivity to be about $41\, mK/\sqrt{Hz}$ at 450\,K. Fig. \ref{fig:5}(c) shows that the temperature sensitivity was maximum at a temperature range from 400\,K to 500\,K. This should be ascribed to the increase of dD/dT at high temperature. However, the decrease of ODMR contrast and fluorescence intensity results in the low sensitivity at higher temperature. As mentioned before, the sensitivity can be improved by increasing coherence time or NV concentration. To obtain a longer coherence time, we can choose $^{12}C$-purified diamond \cite{25} or utilize dynamical decoupling technology \cite{26,27}. Since only one fourth of the NV center in our CVD diamond was used, the employment of preferred-aligned NV- ensemble \cite{28,29} will greatly improve the ODMR contrast, as well as the sensitivity. 

\section{Conclusion}

In conclusion, we have determined the temperature dependence of coherence properties of NV ensemble in diamond from 300\,K to 600\,k. The results reveal that the inhomogeneous dephasing time $T_2^*$ and thermal echo decoherence time $T_{TE}$ are robust to temperature changing. Take advantage of this robust behavior, a TE-based thermometer exhibits a sensitivity of $41\, mK/\sqrt{Hz}$ at 450\,K. However, the SQ and DQ relaxation rates increase rapidly as temperature rises, which are both ascribed to the spin-phonon interaction. Moreover, we report the coherence time $T_2$ of NV center ensemble at temperature up to 600\,K for the first time, the result indicates that the paramagnetic impurities will severely destroy the coherence of NV center. We believe that the investigation of the high temperature coherence properties of NV center will not only be beneficial to broadening the application of NV based devices, but also help to have a more sophisticated understanding of the decoherence in other solid qubits, like divacancy center in silicon carbide (SiC) \cite{30}.

\begin{acknowledgments}
	This work wassupported by the National Basic Research Program of China(Contracts No.\ 2011CB921400 and No.\ 2013CB921800) and the National Natural Science Foundation of China (Grants No.11374280 and No. 50772110). This work was partially carried out at the USTC Center for Micro and Nanoscale Research and Fabrication.
\end{acknowledgments}

\bibliography{bibliography}

\end{document}